\documentclass[aps,twocolumn,floatfix,showpacs]{revtex4}
\usepackage{graphicx,bm}
\usepackage{amsmath}
\usepackage{amssymb}
\begin{document}
\title {Universal 1+2-body bound states in planar atomic wave guides}
\author{Ludovic Pricoupenko$^{(1)}$ and Paolo Pedri$^{(1,2)}$}
\affiliation
{$^{(1)}$ Laboratoire de Physique Th\'{e}orique de la Mati\`{e}re Condens\'{e}e, 
Universit\'{e} Pierre et Marie Curie,\\
4 place Jussieu, 75252 Paris Cedex  05, France.\\
$^{(2)}$ Laboratoire de Physique des Lasers, CNRS-Universit{\'e} Paris 13,\\
99 avenue Jean-Baptiste Cl{\'e}ment, 93430 Villetaneuse, France.
}
\date{\today}
\begin{abstract}
Shallow heteronuclear trimers are predicted for mixtures of two atomic species strongly trapped in a quasi two-dimensional (2D) atomic wave guide. The binding energies are functions of the 2D-scattering length and of the mass ratio and can be thus tuned by various ways. These universal trimers are composed of two identical non interacting particles and of a third particle of the other species. Depending on the statistics of the two identical particles, the trimers have an odd (fermions) or even (bosons) internal angular momentum. These results permit one to draw conclusions on the stability issue for the quasi-2D gaseous phase of heteronuclear dimers.
\end{abstract}
\pacs{03.75.Ss,03.65.Nk,03.65.Ge,05.30.Jp}
%
%
\maketitle

\section{INTRODUCTION}

Actual experiments in ultracold physics open the possibility of trapping and cooling atomic mixtures composed of different atomic species \cite{Ferrari,Roati,Wille,Taglieber} in quasi two-dimensional (quasi-2D) geometries \cite{Rudi_2D,Jean1}. This dramatic progress opens exciting possibilities for exploring experimentally few- and many-body properties in these systems. For example, the Berezinski-Kosterlitz-Thouless transition in bosonic 2D gases is under active experimental studies \cite{Jean2,Jean3,Zoran}. The occurence of confinement induced resonances \cite{Petrov2D_a,Petrov2D_b,LOWD} \emph{i.e.} for quasi-2D wave guide, resonances in two-body scattering induced by the transverse confinement, is another peculiar property in these systems which represents a way to achieve strongly correlated regimes. There are also promising possibilities using fermionic quasi-2D gas to reach exotic phases \cite{Gurarie}. More recently, a new crystalline order has been predicted to occur for two-species quasi-2D fermionic gases with a sufficiently large mass ratio \cite{Petrov_crystal}. This last reference suggests also the idea of tuning the mass ratio by imposing an optical lattice acting on one species only. This opens fascinating possibilities for achieving systems with very high mass ratio. 

Motivated by these recent progress, we consider in this paper mixtures of two atomic species ``strongly confined'' along the $z$-direction by a planar atomic wave guide of frequency $\omega_z$, without any external confinement along the $x$ and $y$ directions. Here, the expression ``strongly confined'' means that for large interparticle separations (\emph{i.e.} for distances which are very much larger than the transverse harmonic oscillator length denoted {$a_z$}), atoms are frozen in the ground state of the 1D harmonic oscillator along $z$. In this situation, the system can be considered quasi-2D and long range properties (for interparticle distances very much larger than $a_z$) acquire the peculiar aspects typical of a 2D-world. The results of the paper can be summarized as follows: (i) We extend the formulation of the 2D zero-range potential (ZRP) approach which is usually used in the configurational space \cite{Lim80,Nielsen,Kar06,Vlambda,HFB2D,Stab2D} to the momentum representation which is more natural for a translation invariant problem. This way the equivalence between the ZRP model and the diagrammatic \cite{Leyronas} or the $t$-matrix approaches \cite{Adhikari} is much more transparent. (ii) We predict the existence of ``${1+2}$-body'' universal shallow bound states composed of two identical atoms of mass $M$ and of a  third atom: an ``impurity'' of mass $m_i$. There is no interaction between the two identical atoms of mass $M$ while each one interacts (in the $s$-wave channel only) with the impurity. For each statistics, the bound states are labeled by a radial quantum number and an internal angular momentum. This latter one is even when identical atoms are bosons and odd when they are fermions. The binding energies are function of the two-body scattering length, of the oscillator length and of the mass ratio, thus they can be tuned by various ways. If the two identical atoms obey the Fermi statistics, we find that for increasing values of the mass ratio ${x=M/m_i}$, the first trimer appears at ${x= x_{\rm 1} \simeq 3.33}$ for a $p$-wave internal angular momentum. (iii) Neglecting inelastic processes toward deep molecular bound states with spatial extensions of the order of the range of interatomic forces (which is very much smaller than ${a_z}$), the low energy two-body collisions between 2D shallow heteronuclear dimers composed of fermions are elastic  for a mass ratio ${x < x_{\rm 2} \simeq 18.3}$. ($iv$) In the case where identical atoms are bosons, at least one trimer exists for all values of the mass ratio. Moreover, binary collisions for dimers composed at least of one boson can always populate a shallow trimer state. 

The structure of the paper is as follows. In Sec. II, we recall the low energy two-body scattering properties for atoms strongly confined in a 2D plane by a harmonic trap along the third dimension. We then introduce the ZRP model in the momentum representation, where the contact condition for a vanishing distance between interacting particles is replaced by an integral equation. In the Sec. III, we explain how this formulation can be extended to few-body problems. In Sec. IV, we show that depending on the various parameters (statistics, mass ratio, 2D-scattering length) the three-body wave equation can support one or more bound-states. We then discuss about the implications of these findings. In particular we focus on the stability issue for the quasi-2D heteronuclear dimer gas.

\section{Momentum representation of the zero-range potential Model}

In this section,  we introduce the ZRP model in the momentum space. We then consider a system composed of two interacting particles of respective masses $M$ and $m_i$ (with the reduced mass $\mu$) interacting in the $s$-wave channel only. The particles are trapped by a harmonic  potential along the $z$-axis with the same trap frequency ($\omega_z$) for both species. We then define the harmonic oscillator length as:
\begin{equation}
a_z=\sqrt{\frac{\hbar}{\omega_z\mu}} .
\end{equation}
According to the ``strong 2D confinement'' hypothesis, we suppose that for large interparticle separations (very much larger than $a_z$), the wavefunction factorizes in the product of a wavefunction of the 2D coordinates (2D-wave function) and of two ground state eigenfunctions of the 1D-harmonic trap for each atomic $z$-coordinate. In what follows, $a_z$ represents a short range scale (in the 2D-world, this length is formally equivalent to the range of the true interatomic forces in three dimensions) and we consider only the quasi-2D behavior: atomic momenta and positions are in the 2D $xy$ plane. The two-body scattering states for an incident plane wave of wave vector ${\bf k}_0$ are given in the low energy regime (${k_0 a_z \ll 1}$) by:
\begin{equation}
\langle {\bf k} | \Psi_{{\bf k_0}} \rangle = (2\pi)^2 \delta({\bf k} - {\bf k}_0) + \frac{2 \pi f_{\rm 2D}(k_0)}{k^2 - k_0^2 -i 0^+}  
\label{eq:2Dscatt} ,
\end{equation}
where $f_{\rm 2D}(k_0)$ is the 2D scattering amplitude \cite{modif_F2D}:
\begin{equation}
f_{\rm 2D}(k_0) = \frac{1}{\ln\left(e^\gamma a_{\rm 2D} k_0/2 \right)-\frac{i \pi}{2}}.
\label{eq:f2D}
\end{equation}
In Eq.~(\ref{eq:f2D}), $\gamma=0.577\cdots$ is the Euler's constant and $a_{\rm 2D}$ is the so-called 2D-scattering length. For a pairwise interaction in 3D characterized by a scattering length $a_{\rm 3D}$, $a_{\rm 2D}$ is a function of the characteristic length of the trap $a_z$ and also of the three dimensional scattering length $a_{\rm 3D}$ \cite{Petrov2D_b}:
\begin{equation} 
a_{\rm 2D}  = \frac{a_z}{\sqrt{e^\gamma \exp(J)}} \exp\left(-\frac{a_z\sqrt{\pi}}{2 a_{\rm 3D}}\right) \, \label{eq:a2D}
\end{equation}
and the constant $J$ in Eq.~\eqref{eq:a2D} is \cite{LOWD,Pri10b}:
\begin{equation}
J =  \int_0^\infty  \frac{du}{u} \left( \frac{1}{\sqrt{1-e^{-u}}}-\frac{1}{\sqrt{u}}-\frac{u}{1+u} \right) = -1.36 \cdots
\end{equation}
Eqs.(\ref{eq:2Dscatt}, \ref{eq:f2D}) can be deduced by using a 2D ZRP model, which by construction is valid for large interparticle spacing with respect to $a_z$ \cite{Vlambda,HFB2D,Stab2D}. Following the spirit of the Bethe-Peierls approach \cite{Bethe}, we recall briefly the formalism in the configurational space for a stationary state ${| \Psi \rangle}$ of energy $E$ in the center of mass frame. The pairwise interaction in the non-interacting Schr\"{o}dinger equation is replaced by a ${\delta({\bf r})}$ source term characterized by an amplitude denoted in what follows by ${\mathcal R}$:
\begin{equation}
\left( - \frac{\hbar^2}{2 \mu} \Delta_{\mathbf r} - E \right) \langle {\mathbf r} | \Psi \rangle =  
\frac{ \pi \hbar^2 }{\mu} \delta({\bf r}) {\mathcal R} .
\label{eq:r-ZRM}
\end{equation}
Solutions of Eq.~(\ref{eq:r-ZRM}) are linear combination of a singular and of a regular function. For a regular function satisfying the boundary conditions associated with a given situation (for example an incident plane wave in the case of a scattering problem), the interacting wavefunction is obtained by determining the balance between these two functions \emph{i.e.} by choosing a specific value for the source amplitude ${\mathcal R}$. In the present case and analogously to what happens for point-charge in 2D-electrostatic, the source term leads to a logarithmic singularity for vanishing interparticle separations ${r \to 0}$. The correct balance between the singular and the regular contributions is deduced by imposing the following behavior on the wave-function as ${{\mathbf r} \to 0}$ \cite{Lim80,Stab2D}:
\begin{equation}
\langle {\bf r}|\Psi \rangle =-{\mathcal R} \ln\left(\frac{r}{a_{\rm 2D}}\right) + O(r) .
\label{eq:r-contact_2D}
\end{equation}
This contact condition allows one to recover the scattering states in Eq.~(\ref{eq:2Dscatt}) exactly. It is worth pointing out that the logarithmic singularity is purely formal because for $r$ of the order or smaller than $a_z$, the 2D zero range model ceases to be valid and 3D physics becomes important. However, this is the key property which permits to implement the effect of the interaction for interparticle separations very much larger than $a_z$. The systems which are studied in this paper are translationally invariant along the $xy$ plane. Hence, solving the stationnary Schr\"{o}dinger equation is a much more simple task in the ${\mathbf k}$-representation than in the configurational space. Consequently, we formulate below the 2D ZRP model in the momentum representation. For this purpose, we closely follow the method already introduced in Ref.~\cite{LOWD} for arbitrary resonant $l$-wave interaction in 3D. First, we introduce a source term with a delta of finite range ${\epsilon}$:
\begin{equation}
\langle {\mathbf k} |\delta_\epsilon\rangle= \exp\left(-\frac{k^2\epsilon^2}{4}\right) ,
\end{equation}
which is such that: ${\lim_{\epsilon \to 0} \langle {\mathbf r} |\delta_\epsilon\rangle=\delta ({\mathbf r})}$. Keeping the same notations as in Eq.~(\ref{eq:r-ZRM}), the stationary states $|\Psi_\epsilon\rangle$ at energy $E$ verify:
\begin{equation}
\left( \frac{\hbar^2 k^2}{2\mu} - E \right) \langle {\bf k} | \Psi_\epsilon \rangle
= \frac{ \pi \hbar^2 }{\mu} \langle {\mathbf k} |\delta_\epsilon\rangle  {\mathcal R}.
\label{eq:2bodyZRM}
\end{equation}
and the stationary states in the zero-range limit are:
\begin{equation}
{|\Psi \rangle = \lim_{\epsilon \to 0} |\Psi_\epsilon \rangle} .
\end{equation}
Let us consider the 2D Green's function of Eq.~\eqref{eq:r-ZRM} (solution corresponding to $\mathcal R=1$) taken at energy $-\hbar^2\Lambda^2/\mu<0$ where $\Lambda$  is any real and positive  number. This function is $K_0(\Lambda r)$ and behaves as $-\ln\left(\Lambda r e^\gamma/2\right) +O(r)$ for $r\to0$. Using this property, the contact condition in Eq.~(\ref{eq:r-contact_2D}) can be written in terms of a subtraction between the eigenstate of the ZRP model and this Green's function as $r\to 0$ \cite{Pri10b}. In the ${\mathbf k}$-representation the contact condition can thus be written as
\begin{equation}
\lim_{\epsilon \to 0} \int \frac{d^2{\bf k}}{(2\pi)^2} 
\left(\langle {\bf k} | \Psi_\epsilon  \rangle 
-\frac{2 \pi\langle {\mathbf k} |\delta_\epsilon\rangle  {\mathcal R} }{k^2 +\Lambda^2} \right) 
= \frac{{\mathcal R}}{f_{\rm 2D}(i \Lambda )} ,
\label{eq:contact}
\end{equation}
where the scattering amplitude at negative energy is:
\begin{equation}
f_{\rm 2D}(i \Lambda ) = \frac{1}{\ln\left(e^\gamma a_{\rm 2D} \Lambda/2\right)}  \qquad \Lambda >0.
\end{equation}
We emphasize that the interacting wave function obtained by using Eq.~(\ref{eq:contact}) do not depend on the parameter $\Lambda$: this is the  so-called $\Lambda$-freedom introduced previously through the $\Lambda$-potential in Ref.~\cite{Vlambda}. As an example, we illustrate this method in the case of a two-body collision for an incident plane wave of momentum ${{\mathbf k}_0}$ in the center of mass frame. For ${E>0}$, using the usual prescription ${E \to E  + i 0^+}$ corresponding to an outgoing cylindrical wave, Eq.~(\ref{eq:2bodyZRM}) can be transformed into:
\begin{equation}
\langle {\bf k} | \Psi_\epsilon \rangle  =  (2\pi)^2 \delta({\bf k} - {\bf k}_0)  + \frac{ \pi \hbar^2 }{\mu} 
\frac{ \langle {\mathbf k} |\delta_\epsilon\rangle{\mathcal R}}{\frac{\hbar^2 k^2}{2\mu} -E -i0^+}.
\label{eq:ansatz}
\end{equation}
Then, inserting Eq.~(\ref{eq:ansatz}) in the contact condition Eq.~(\ref{eq:contact}), one obtains  ${{\mathcal R} = f_{\rm 2D}(k_0)}$ given by Eq.~(\ref{eq:f2D}) ${\forall \Lambda>0}$. Therefore as expected, performing the $\epsilon \to 0$ limit in Eq.~(\ref{eq:ansatz}) one recovers the 2D $s$-wave scattering states of Eq.~(\ref{eq:2Dscatt}). The 2D ZRP model supports also the existence of a two-body bound state characterized by a binding energy ${E_{\rm dim}=-E}$ where
\begin{equation}
E_{\rm dim} = \frac{\hbar^2 q_{\rm dim}^2 }{2\mu} \quad \mbox{with} \quad  q_{\rm dim}=\frac{2}{e^\gamma a_{\rm 2D}} ,
\label{eq:Edim}
\end{equation}
and a wavefunction $|\phi_B\rangle$ given in the ${\mathbf r}$-representation by
\begin{equation}
\langle {\mathbf r} |\phi_B\rangle = \frac{q_{\rm dim}}{\sqrt{\pi}} K_0(q_{\rm dim} r)
\end{equation}
This bound state has a spatial extension of the order of $a_{\rm 2D}$ and can thus be considered as shallow in the limit
\begin{equation}
a_{\rm 2D} \gg a_z .
\label{eq:shallow-regime}
\end{equation}
Equation~\eqref{eq:shallow-regime} defines the quasi-2D resonant regime where the extension of the 2D dimer is large with respect to $a_z$ and is analogous to the 3D resonant regime where the 3D scattering length is large in absolute value with respect to the typical radius of interatomic forces. The length $a_z$ gives the order of magnitude of the radius of the 2D effective interaction \cite{Petrov2D_b} so that the quasi-2D resonant regime defines also the regime where the zero-range approximation is well justified. For instance, the scattering amplitude in Eq.~\eqref{eq:f2D} has a maximum modulus for ${k=q_{\rm dim}}$ and in the quasi-2D resonant regime, the maximum occurs for a 2D momentum which is very small with respect to ${1/a_z}$ \emph{i.e.} for a collisional energy such that the 2D approximation remains valid. For a fixed value of the collisional momentum, the maximum can be reached if one tunes the dimer binding wavenumber ${q_{\rm dim}}$ (and thus the 2D scattering length) by modifying the 3D-scattering length and/or the trap frequency $\omega_z$ [see Eq.~\eqref{eq:a2D}]: this is the {\it so-called} 2D Confinement Induced Resonance first found in Ref.~\cite{Petrov2D_a}. As a consequence of Eq.~\eqref{eq:a2D}, the condition in Eq.~(\ref{eq:shallow-regime}) corresponds to the regime of negative 3D-scattering lengths where also ${|a_{\rm 3D}|}$ is much smaller than $a_z$. In this paper, we suppose that Eq.~\eqref{eq:shallow-regime} is satisfied for heteronuclear atomic pairs.

Note that in the case where an optical lattice in the $xy$-plane acts for example on the heavy atoms of mass $M$ in the small filling limit as proposed in Ref.~\cite{Petrov_crystal}, we assume that $a_{\rm 2D}$ is much larger than the lattice period. In this limit the mass $M$ can be replaced by the effective mass in the 2D-Schr\"odinger equation \cite{Fedichev}.

\section{Formulation of the ${1+2}$-body problem}

In this second part, we show how the ZRP model can be implemented in few-body systems and in the momentum representation by considering a particular '${1+2}$-body' problem. Our system is composed of two identical atoms of mass $M$ each interacting only with a single atom impurity of mass $m_i$  via the ZRP model. Hence in order to fulfill the strong 2D confinement condition's, we require implicitly that the collisional energies of the problem are very much smaller than the characteristic energy of the transverse trap ($\hbar\omega_z$).

Furthermore, we assume that the quasi-2D resonant regime defined in Eq.~(\ref{eq:shallow-regime}) is always achieved for heteronuclear atomic pairs. We introduce the statistical parameter $\eta$ which is equal to $-1$ if the two identical atoms are fermions, and to $+1$ if they are bosons. In the fermionic case, the two identical atoms do not interact in the $s$-wave channel as a consequence of the Pauli principle. In the bosonic case, we also consider that the identical atoms do not interact in the sense that their pairwise interaction is negligible with respect to the interaction with the other atomic species. This hypothesis is justified in the case where the 3D scattering length between the two identical bosons is positive and small in comparison with $a_z$ which implies that their 2D scattering length is exponentially small in comparison to ${a_z}$ [see Eq.~\eqref{eq:a2D}]. Consequently, in both cases we take into account the pairwise interaction only between particles of different mass. In the ZRP model, for each interacting pair ${(i,j)}$, the interaction is implemented by inserting in the Schr\"{o}dinger equation a source term with an amplitude denoted by ${{\mathcal R}^{i \leftrightharpoons j}}$. Consequently, we assume implicitly that configurations where three particles are in a hyper-radius smaller or of the order of $a_z$ can be neglected. This hypothesis is especially justified in the quasi-2D resonant regime where the characteristic extension of the trimers (and thus of the hyper-radius) is of the order of ${a_{\rm 2D}}$ which is very large with respect to ${a_z}$. In the following, we associate the three momenta $\{{\bf k}_i\} \equiv \{{\bf k}_1,{\bf k}_2,{\bf k}_3 \}$ with the particle configuration: ${(1 :\, M \, ;\, 2 :\, M\,; 3 :\, m_i \, )}$, furthermore we solve the problem in the center of mass frame so that:
\begin{equation}
{\bf k}_1+{\bf  k}_2+{\bf k}_3={\bf 0} .
\label{eq:Cdm}
\end{equation}
For each pair $(i,j)$ the relative coordinates denoted by ${\boldsymbol \eta}_l$ (where $i,j,l$ are in cyclic order) verify:
\begin{equation}
{\boldsymbol \eta}_1=\frac{{\bf  k}_2-x{\bf k}_3}{1+x} \quad
{\boldsymbol \eta}_2=\frac{x{\bf  k}_3-{\bf k}_1}{1+x} \quad
{\boldsymbol \eta}_3=\frac{{\bf k}_1-{\bf k}_2}{2} ,
\end{equation}
where ${x}$ is the mass ratio ${x=M/m_i}$. Using these notations, the Schr\"{o}dinger equation for an eigenstate ${|\Psi_\epsilon \rangle}$ of energy $E$ reads in the momentum representation:
\begin{eqnarray}
&& \left[ \frac{\hbar^2  (k_1^2+k_2^2)}{2M}+ \frac{\hbar^2k_3^2}{2m_i}  -  E \right] 
\langle \{ {\bf k}_i \} |\Psi_\epsilon \rangle =  \frac{ \pi \hbar^2 }{\mu}  \big[
\nonumber \\
&& \qquad   {\mathcal R}^{1\leftrightharpoons 3} \langle {\boldsymbol \eta}_2  |\delta_\epsilon\rangle 
+  {\mathcal R}^{2 \leftrightharpoons 3} \langle {\boldsymbol \eta}_1  |\delta_\epsilon\rangle \big] .
\label{eq:Schrodi1}
\end{eqnarray}
In Eq.~(\ref{eq:Schrodi1}), ${{\mathcal R}^{i \leftrightharpoons j}}$ is a function of the total momentum of the pair ${(i,j)}$ and of the momentum of the third particle ($l$). Using the fact that we solve the problem in the center of mass frame, ${{\mathcal R}^{i \leftrightharpoons j}}$ can be considered as a function of the momentum ${{\mathbf k}_l}$. Moreover, the statistics of the two identical particles imposes a symmetry between the two source terms:
$ {\mathcal R}^{1\leftrightharpoons 3}({\mathbf k})  = \eta  {\mathcal R}^{2\leftrightharpoons 3}({\mathbf k})$. In order to simplify the subsequent equations, we will denote the amplitude ${\mathcal R}^{2\leftrightharpoons 3}({\mathbf k})$ by ${\mathcal R}({\mathbf k})$. We also consider in what follows situations of negative total energy $E$ only, we thus introduce the momentum $q$:
\begin{equation}
E= -\frac{\hbar^2 q^2}{2 \mu}  \quad \mbox{where} \quad q>0 .
\end{equation}
The equation satisfied by the function ${{\mathcal R}({\bf k})}$ is deduced from the contact condition applied to an interacting pair. For example, application of Eq.~(\ref{eq:contact}) for the pair ${(2 : M \,; 3 :\, m_i \, )}$ of relative momentum ${{\boldsymbol \eta}_1}$ gives:
\begin{equation}
\int \frac{d^2 {\boldsymbol \eta}_1} {(2\pi)^2}\, \left[ \langle \{ {\bf k}_i \} | \Psi \rangle 
- \frac{2 \pi {\mathcal R} ({\bf k}_1)}{{\boldsymbol \eta}_1^2 + \Lambda^2}\right] = \frac{{\mathcal R} ({\bf k}_1)} {f_{\rm 2D}(i\Lambda)} .
\label{eq:contact_S} 
\end{equation}
For ${E<0}$, Eq.~(\ref{eq:contact_S}) can be simplified as:
\begin{equation}
\frac{{\mathcal R}({\bf k})}{f_{\rm 2D}(\kappa_k)}=
\eta \int \frac{d^2{\bf u}}{(2\pi)^2}\, \frac{2\pi {\mathcal R}({\bf u})}{u^2 +k^2 + 2 y {\bf k}.{\bf u} +q^2} ,
\label{eq:threebody}
\end{equation}
where the variable ${y}$ is a function of the mass ratio: 
\begin{equation}
y=\frac{M}{M + m_i} = \frac{x}{1+x} ,
\end{equation}
and also we have introduced the collisional momentum:
\begin{equation} 
\kappa_{{\bf k}} = i \sqrt{(1-y^2) k^2 + q^2} .
\label{eq:kappa_rel}
\end{equation}
The physical interpretation of Eq.~(\ref{eq:kappa_rel}) proceeds as follows. The scattering process between the two interacting particles (${2,3}$) occurs in their center of mass frame at a kinetic energy ${E^{\rm (2,3)}_{\rm col}}$ which verifies:
\begin{equation}
E = E^{\rm (2,3)}_{\rm col} +  \frac{\hbar^2 ( {\mathbf k}_2 + {\mathbf k}_3 )^2 }{2 (M + m_i)} 
+  \frac{\hbar^2}{2 M} {\mathbf k}_1^2 .
\label{eq:def_Erel}
\end{equation}
In Eq.~(\ref{eq:def_Erel}) we have used the fact that, as mentioned previously, the third atom ($1$ in this case) does not interact with an atom of the pair $(2,3)$ during the scattering process. Hence, using Eqs.(\ref{eq:kappa_rel},\ref{eq:def_Erel}) the collisional energy of the pair can be written as:
\begin{equation}
E^{\rm (2,3)}_{\rm col} =\frac{\hbar^2 \kappa_{{\bf  k}_1}^2 }{2\mu}  < 0 ,
\end{equation}
and ${\kappa_{{\bf  k}_1}}$ is the relative momentum of the pair (${2,3}$) of total momentum ${(-{\mathbf k}_1)}$. Eq.~\eqref{eq:threebody} is the 2D analog for the $1+2$-body problem of the so-called Skorniakov Ter-Martirosian equation \cite{Sko57}. For three identical bosons in 2D where ${{\mathcal R}^{1\leftrightharpoons 2}={\mathcal R}^{1\leftrightharpoons 3}={\mathcal R}^{2\leftrightharpoons 3}}$, a similar eigenequation has been already derived in \cite{Bruch,Adhikari,Leyronas}
(in that case, ${\eta=2}$ and ${y=1/2}$ in Eq.~\eqref{eq:threebody}). An analogous integral equation has been obtained in the context of quasi-2D heteronuclear trimers in Ref.~\cite{Lev09} where the third direction is also taken into account.

\section{$\mathbf{1+2}$-body bound states}

In this last part, we show that Eq.~(\ref{eq:threebody}) supports the existence of trimers which can be considered as shallow in the regime defined by Eq.~(\ref{eq:shallow-regime}). We denote their binding energy by ${E_{\rm trim}}$:
\begin{equation}
E=-E_{\rm trim} < - E_{\rm dim} .
\label{eq:Etrim}
\end{equation}
The kernel in Eq.~(\ref{eq:threebody}) has the cylindrical symmetry, hence trimers states can be labeled by a radial quantum number and an orbital quantum number denoted below by $m$. We thus expand the source amplitude  on the $m$-partial waves:
${{\mathcal R}({\bf k })= \sum_{m=0}^\infty \cos(m\theta) {\mathcal R}_m(k)}$, with ${\theta = \angle ({\bf \hat{e}}_x,{\bf k })}$.  The angular integration of the kernel can be performed for each $m$-partial wave, and one obtains the following integral equation:
\begin{eqnarray}
\frac{{\mathcal R}_m(k)}{f_{\rm 2D}(\kappa_{\mathbf k})}  =  
\frac{\eta (-1)^{m}}{2ky} \int_0^\infty du \, \frac{ \left( t - \sqrt{t^2 - 1}  \right)^m}{\sqrt{t^2 - 1} }  {\mathcal R}_m(u)  ,
\label{eq:threebody_m}
\end{eqnarray}
where ${t= {(u^2+k^2+q^2)}/{(2y k u)} > 1}$. Eqs.(\ref{eq:kappa_rel},\ref{eq:Etrim}) imply that ${f_{\rm 2D}(\kappa_{\mathbf k})}$ is positive. Whence, for $\eta=+1$ (atoms of mass $M$ are bosons) Eq.~(\ref{eq:threebody_m}) admits solutions for $m$ even only, while for $\eta=-1$ (identical atoms are fermions) the orbital quantum number of the trimers is necessarily odd. We have computed the spectrum for mass ratios $x$ up to $200$ -such high values can be reached experimentally using an optical lattice \cite{Petrov_crystal}.

\begin{figure}[h]
\includegraphics[width=8cm,clip]{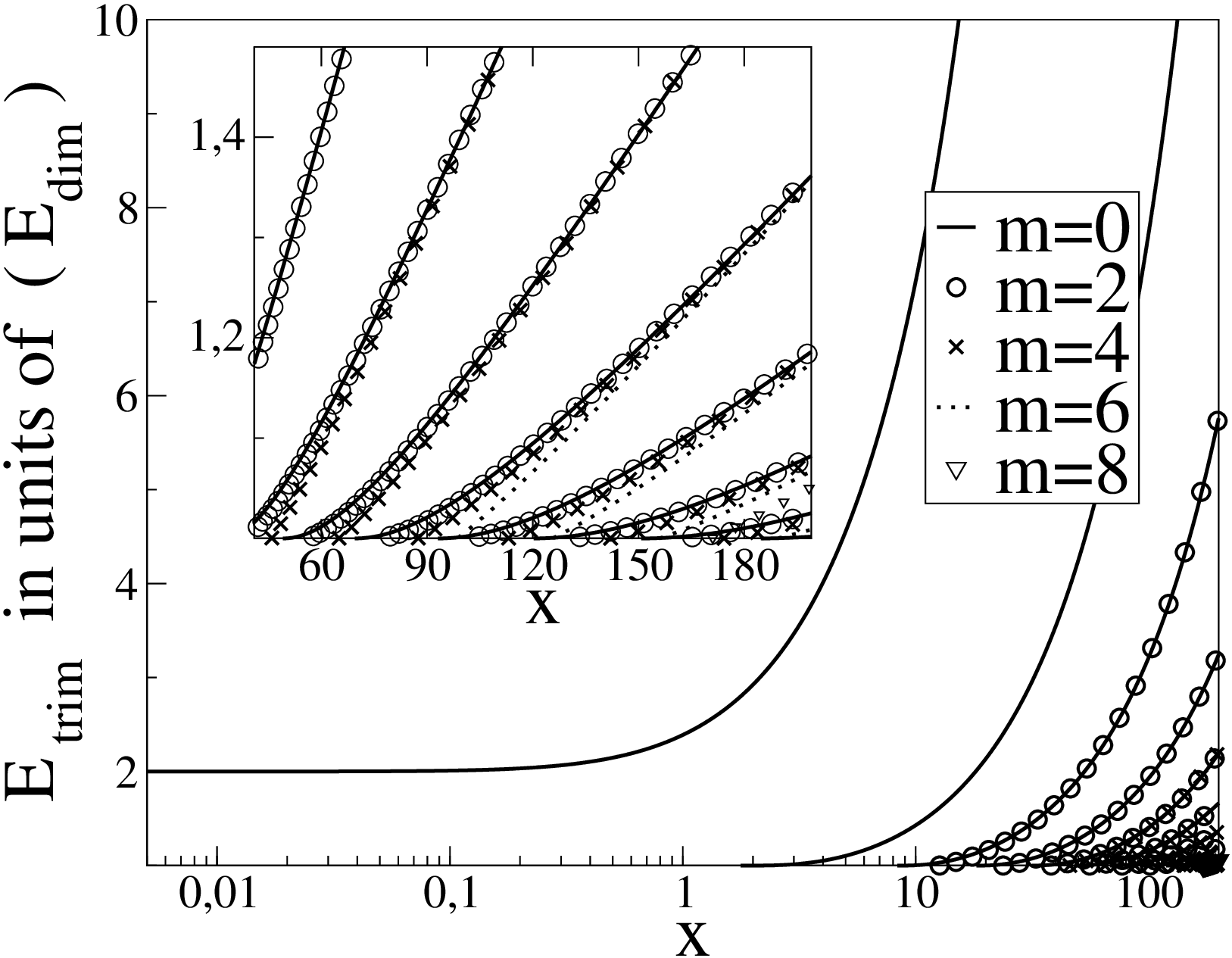}
\caption{Trimers' spectrum for two identical bosons of mass $M$ interacting with a third atom of mass $m_i$. The binding energies ${E_{\rm trim}}$ are expressed in unit of the heteronuclear dimer's energy $E_{\rm dim}$ in Eq.~(\ref{eq:Edim}), as a function of the mass ratio ${x=M/m_i}$. One can notice that away from threshold, the bound state energy of a $m$-trimer is quasi-degenerate with the energy of a trimer with a lower $m$.}
\label{fig:trimers_B}
\end{figure}

{\it Bosonic spectrum}-- The bosonic spectrum is plotted in Fig.~(\ref{fig:trimers_B}). For $x \le 200$, we have found trimers with ${m\le 8}$ \cite{Trim3bosons} and the deepest trimer is in the $m=0$ sector. Apparition threshold of three first excited trimers in the $m=0$ sector are located at $x=1.77$, $x=8.34$ and $x=18.27$. Trimers of increasing orbital momentum appear as $x$ increases: as an example, $m=2$ trimers appear for $x \gtrsim 12.68$. An important feature of the bosonic spectrum is that for all possible values of the mass ratio, there exists at least one trimer state. In the specific case where ${m_i=M}$ and $m=0$, we recover the result of Ref.~\cite{Leyronas}: there is only one bound state with a binding energy at $E_3/E_{\rm dim}=2.36\cdots$ 

\begin{figure}[h]
\includegraphics[width=8cm,clip]{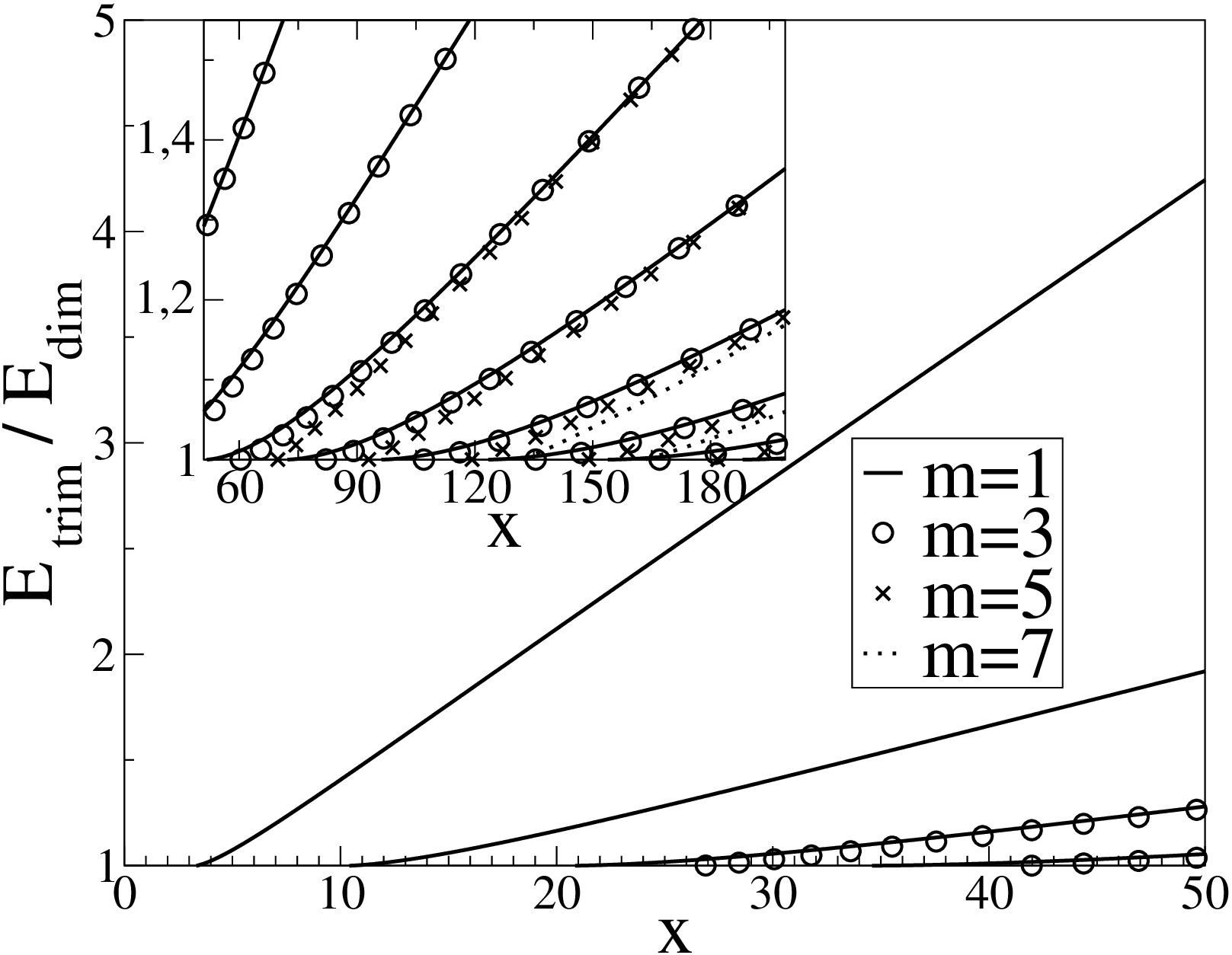}
\caption{Same as Fig.~(\ref{fig:trimers_B}) for two identical fermions of mass $M$ interacting with a third atom of mass ${m_i}$.}
\label{fig:trimers_F}
\end{figure}

{\it Fermionic spectrum}-- The fermionic spectrum corresponds to the case where the two identical atoms are fermions (${\eta=-1}$) and is plotted in Fig.~(\ref{fig:trimers_F}).
We find that the deepest shallow trimer has a $p$-wave internal angular momentum (${m=1}$) and exists only for  ${x>x_{\rm 1}\simeq 3.33}$. For increasing values of ${x>x_{\rm 1}}$, other shallow trimer states appear in the ${m=1}$ sector and for higher odd values of $m$.  

Numerical values of mass ratios for the threshold of apparition of ${1+2}$-trimers in the $m\le8$ momentum sectors are gathered in Tabs.~(\ref{tab:I},\ref{tab:II}).

\begin{table}[hx]
\begin{tabular}{|c|c|c|c|c|c|c|c|}
\hline
        & ${m=0}$   & ${m=1}$ 	& ${m=2}$	&${m=3}$	& ${m=4}$ 	& ${m=5}$ 	& ${m=6}$\\ \hline
${n=0}$ &   -       &	$3.340$	& $12.68$    	&$26.89$	& $45.95$	&$69.83$	&  $98.84$\\\hline
${n=1}$ &  $1.770$   &	$10.41$	& $23.76$    	&$41.98$	& $65.02$	&$92.97$	&  $125.8$\\\hline
${n=2}$ &  $8.341$   &	$20.85$	& $38.17$    	&$60.36$	& $87.49$	&$119.3$	&  $155.9$\\\hline
${n=3}$ &  $18.27$   &	$34.59$	& $55.91$    	&$82.08$	& $113.1$	&$149.0$	&  $189.7$\\\hline
${n=4}$ &  $31.62$   &	$51.76$	& $77.04$	&$107.0$	& $142.09$	&$181.8$	&  \\\hline
${n=5}$ &  $48.33$   &	$72.28$	& $101.5$	&$135.5$	& $174.2$	&		&  \\ \hline
${n=6}$ &  $68.38$   &	$96.23$	& $129.2$	&$167.0$	&		&		&  \\ \hline
${n=7}$ &  $92.01$   &	$123.5$	& $160.1$	&		&		&		&  \\ \hline
${n=8}$ &  $119.0$    &	$153.8$	& $194.8$	&		&		&		&  \\ \hline
${n=9}$ &  $149.4$    &	$188.2$	&		&		&		&		&  \\ \hline
${n=6}$ &  $182.8$    &		&		&		&		&		&  \\ \hline
\end{tabular}
\caption{Threshold values of the mass ratio $M/m_i$ for the apparition of $1+2$-body bound states. The momentum of the states is labeled by $m$ while $n$ is the hyperradial quantum number. Computations have been limited to mass ratios less than $200$.}
\label{tab:I}
\end{table}

\begin{table}[hx]
\begin{tabular}{|c|c|c|}
\hline
        & ${m=7}$	& ${m=8}$ 	\\\hline
${n=0}$ &  $132.3$    	&$171.0$	\\\hline
${n=1}$ &  $163.5$	&	\\\hline
${n=2}$ &  $197.9$	&	\\\hline
\end{tabular}
\caption{Same as in Tab.~(\ref{tab:I}) for higher momentum.}
\label{tab:II}
\end{table}

\begin{figure}[h]
\includegraphics[width=8cm,clip]{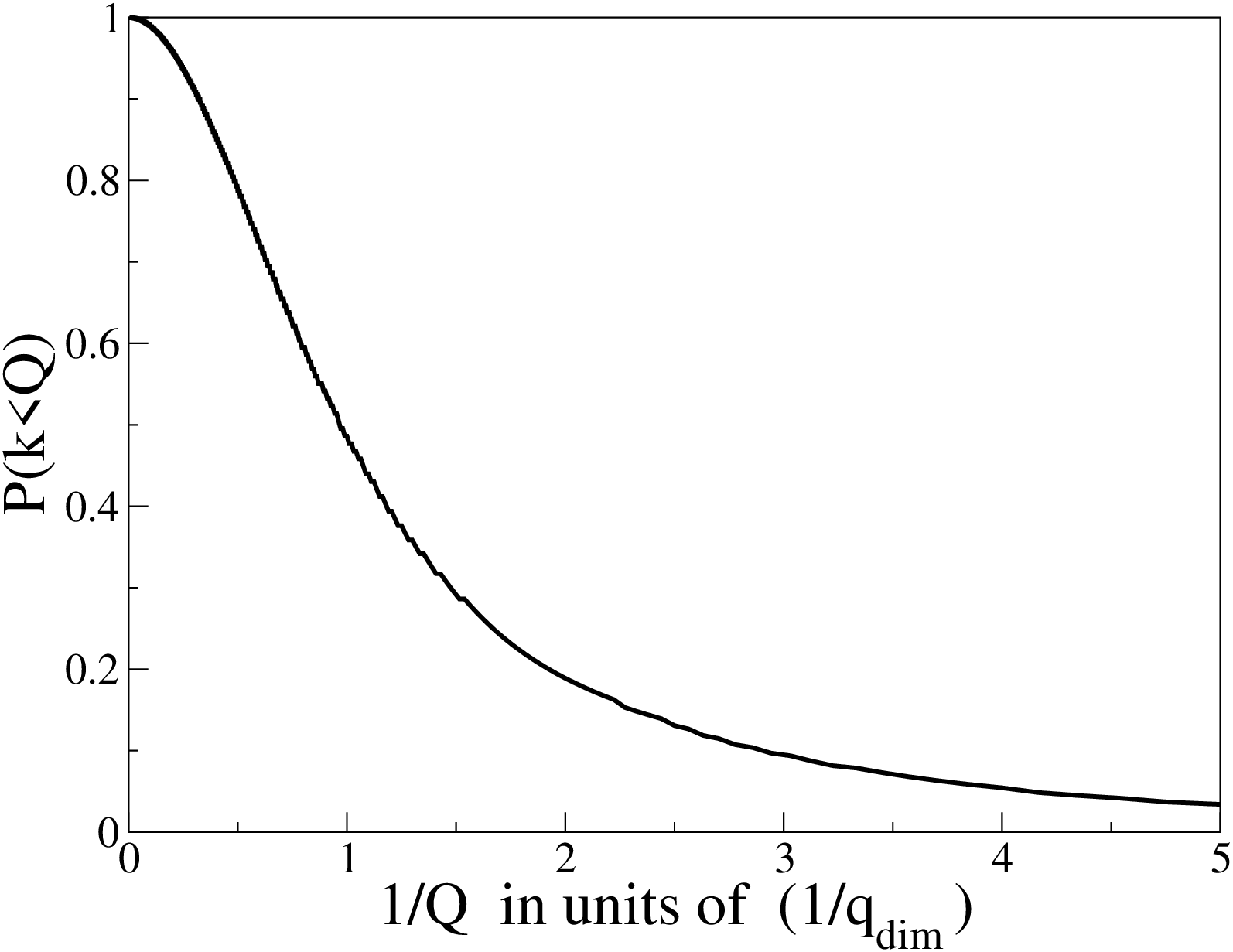}
\caption{Plot of the probability in  Eq.~(\ref{eq:Proba}) ${P(k<Q)}$ as a function of ${1/Q}$ in unit of ${1/q_{\rm dim}}$ for a ${m=1}$-trimer at threshold (${x=153.9}$ and ${q-q_{\rm dim} \ll q_{\rm dim}}$).}
\label{fig:Proba_Q}
\end{figure}

{\it Discussion}-- One important implication of the existence of these universal trimers concerns losses resulting from two-body collisions of heteronuclear shallow dimers. In the case of bosonic identical atoms of mass $M$ [see Fig.~(\ref{fig:trimers_B})] the binding energies are ${E_{\rm trim} \ge 2 E_{\rm dim}}$ for all values of the mass ratio. Thus, collisions of two shallow heteronuclear dimers can always lead to the formation of one ${(MMm_i)}$-trimer plus a single impurity atom of mass ${m_i}$. Therefore, two-body losses prevent the existence of a stable 2D gas composed of such dimers at equilibrium whatever the statistics of the impurity (atoms of mass $m_i$). For fermionic atoms of mass $M$, the deepest trimer has a binding energy ${E_{\rm trim}\ge 2 E_{\rm dim}}$ for mass ratios ${x > x_{\rm 2} \simeq 18.3}$. Thus two-body collisions of heteronuclear dimers where both atomic species are fermions, are elastic for a mass ratio ${x<x_{\rm 2}}$ \cite{Caution}. Whence a gas composed of such shallow dimers is expected to be relatively stable for ${x < x_{\rm 2}}$: it is the case for $^6$Li-$^{40}$K or $^6$Li-$^{87}$Rb mixtures strongly confined in 2D providing Eq.~(\ref{eq:shallow-regime}) is satisfied, while for a $^6$Li-$^{171}$Yb mixture the critical mass ratio is exceeded.
\begin{figure}[h]
\includegraphics[width=8cm,clip]{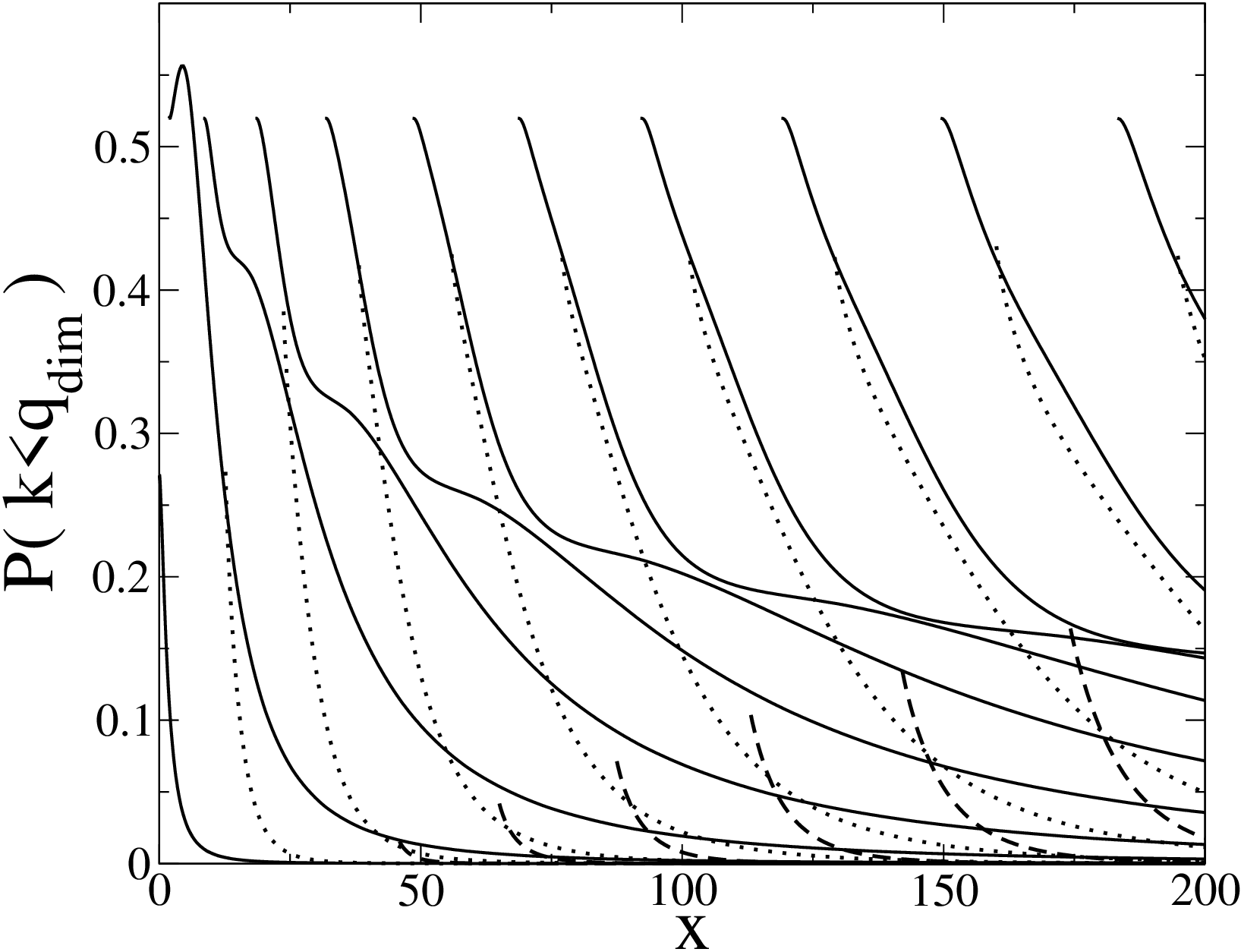}
\caption{Plot of the probability in Eq.~(\ref{eq:Proba}) ${P(k<Q)}$ at ${Q=q_{\rm dim}}$ as a function of the mass ratio $x$ for the different 
trimers composed of bosonic atoms of mass $M$ with the binding energies of Fig.~(\ref{fig:trimers_B}). Continuous lines: ${m=0}$ states, dotted lines: ${m=2}$ states, dashed lines: ${m=4}$ states.}
\label{fig:Proba_B_x}
\end{figure}
In order to get some informations about this inelastic process, we consider for a given normalized trimer state $|\Psi \rangle$, the probability:
\begin{equation}
P(k < Q) = \int _{|{\mathbf k}_1|,|{\mathbf k}_2| < Q}  \frac{d^2{\mathbf k}_1}{(2\pi)^2} \frac{d^2{\mathbf k}_2}{(2\pi)^2} | \langle \{ {\bf k}_i \} | \Psi \rangle |^2 .
\label{eq:Proba}
\end{equation}
${P(k<Q)}$ in Eq.~(\ref{eq:Proba}) gives the order of magnitude that the three particles of the trimer are separated each one from the others by distances greater than or of the order of ${1/Q}$. The results for bosonic atoms of mass ${M}$ and ${Q=q_{\rm dim}}$ is given in 
Fig.~(\ref{fig:Proba_B_x}) and a similar plot for fermions is given in Fig.~(\ref{fig:Proba_F_x}). These figures show clearly that the probability is maximum nearby the formation threshold of a trimer. 
\begin{figure}[h]
\includegraphics[width=8cm,clip]{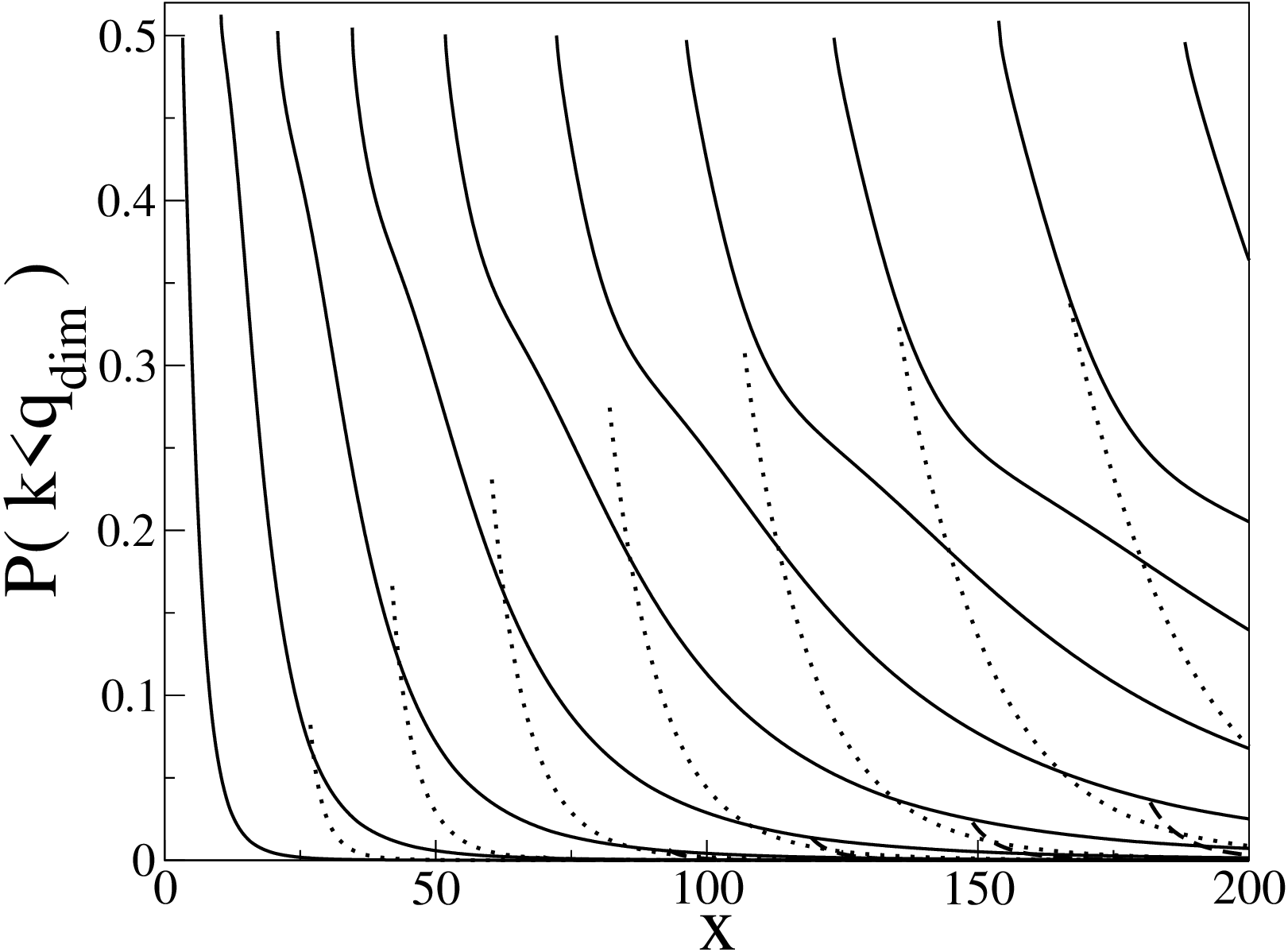}
\caption{Same as Fig.~(\ref{fig:Proba_B_x}) for fermionic atoms of mass $M$ with binding energies of Fig.~(\ref{fig:trimers_F}). Continuous lines: ${m=1}$ states, dotted lines: ${m=3}$ states, dashed lines: ${m=5}$ states.}
\label{fig:Proba_F_x}
\end{figure}
Moreover, the probability at threshold decreases rapidly as the value of the internal angular momentum increases. In Fig.~(\ref{fig:Proba_Q}) we have plot ${P(k<Q)}$ as a function of ${1/Q}$ for a ${m=1}$ trimer near its threshold of apparition (${x=153.9}$): the probability drops out dramatically as one considers increasing values of the interparticle spacing: for ${Q= q_{\rm dim}/2}$, it is still greater than ${10\%}$ while for ${Q = q_{\rm dim}/10}$, it is less than ${1\%}$. Therefore, we can conclude that the spatial extension of the lowest energy trimers is of the order of the dimer's size.
Moreover, it has been shown that the long range effective heteronuclear pairwise interaction is repulsive for high mass ratio with a barrier located at ${\sim 1/q_{\rm dim}}$ \cite{Petrov_crystal}. Thus depending on the statistics, these results show that the dimer-dimer inelatic decays populate predominantly the ${m=0}$ or ${m=1}$ shallow trimers nearby their apparition threshold (and also to less extents the ${m=2}$ and ${m=3}$ trimers). Furthermore, from Figs.(\ref{fig:Proba_B_x},\ref{fig:Proba_F_x}) it appears that such low energy 3-body bound states of size ${\sim 1/q_{\rm dim}}$ exist for all high mass ratios. Hence this motivate future detailed study of such four-body processes which are the most important source of 2D-losses even for high mass ratios.

\section{CONCLUSION}

In this paper, we have derived the low energy three-body wave equation using a ZRP model in the momentum representation for atoms strongly confined in a planar wave guide. In this representation Green's functions are simply rational fractions, as a consequence technical implementation of the formalism is greatly enhanced with respect to the usual configurational space approach. Generalization to others dimensions and other few-body problems can be easily performed~\cite{Pri10b}.

Depending on the statistics and on the mass ratio, we have predicted the existence of quasi-2D heteronuclear trimers composed of two identical atoms  and of one another atom species. For a planar wave guide achieved by using a harmonic trapping along the tight direction (harmonic oscillator length ${a_z}$) these trimers can be effectively considered as quasi-2D in a regime where the 3D heteronuclear scattering length ${a_{\rm 3D}}$ is negative and ${|a_{\rm 3D}|\ll a_z}$. While for fermions the heteronuclear interaction is naturally negligible as a consequence of the Pauli principle, for identical bosons we have considered situations where this channel of interaction is also negligible. Considering the shallow trimers as the most important source of decay in (heteronuclear-dimer)-(heteronuclear-dimer) scattering processes, we have shown that a dimer gas composed of fermionic atoms is stable with respect to binary collisions for a mass ratio ${M/m_i \lesssim 18.3}$ while for bosons the gas is never stable. In the fermionic case, for a fixed value of the areal dimer density $n$ compatible with a 2D strong confinement (\emph{i.e.} $n a_z^2 \ll 1$), knowing that the 2D heteronuclear atomic scattering length can be tuned to an arbitrarily large value for ${a_{\rm 3D}<0}$, then the 2D gas parameter ${na_{2D}^2}$ can reach  high value. This opens the possibility of achieving a stable 2D strongly correlated regime where one can expect also that for a mass ratio ${M/m_i>3.33}$ the shallow trimer state(s) affect(s) the low energy gas properties. Practical achievement of such a low energy dimer gas in harmonic traps requires very high aspect ratios such that ${a_z \ll a_{\rm 2D} \ll a_{\parallel}}$ where $a_{\parallel}$ is the harmonic oscillator length along the direction perpendicular to the tight confinement.

Interestingly, similar results concerning shallow heteronuclear trimers have been found in 3D \cite{Kar07} where the zero-range approximation is also performed and the Shr\"{o}dinger equation is solved within the hyperspherical formalism. In 3D, the critical mass ratio for the first $p$-wave shallow trimer is found at $8.18$ (a larger mass ratio than in 2D). The universal shallow $p$-wave trimers exists also only for positive 3D scattering length (${a_{\rm 3D}>0}$), while in 2D the analogous universal shallow trimer states exist for ${a_{\rm 2D} \gg a_z}$ and thus ${a_{\rm 3D}<0}$. Moreover in Ref.~\cite{Kar09}, exact results have been found for 1D universal heteronuclear trimers. A recent Letter \cite{Lev09} has explored three-body physics in quasi-2D atomic planar waveguide. In this last reference the effective range of the 2-body interaction and also the transverse direction of the wave guide are both taken into account in the three-body problem. Eq.~\eqref{eq:threebody} in our paper appears thus as a 2D limit of this approach.

\begin{center}
{\bf ACKNOWLEDGMENTS}
\end{center}

Oleg Kartavtsev is acknowledged for a careful reading of the manuscript. Laboratoire de Physique Th\'{e}orique de la Mati\`{e}re Condens\'{e}e is Unit\'{e} Mixte de Recherche 7600 du CNRS and its Cold Atoms group is associated with IFRAF. Laboratoire de Physique des Lasers is 'Unit\'{e} Mixte de Recherche 7538 du CNRS' and is a member of IFRAF.


\begin{thebibliography}{99}
\bibitem{modif_F2D} We have taken here a definition of the 2D scattering amplitude which differs from the choice in Ref.~\cite{LOWD}. Using the present convention, the asymptotic scattering states in Eq.~(\ref{eq:2Dscatt}) are given in the ${\mathbf r}$-representation, by: ${\langle {\mathbf r} | \Psi_{{\mathbf k}_0} \rangle = \exp(i{\mathbf k}_0\cdot{\mathbf r}) + \frac{i \pi}{2} f_{\rm 2D}(k_0)  H_0^{(1)}(k_0 r)}$
\bibitem{Trim3bosons} In the case of three identical bosons, we recover the two universal shallow trimers already found in \cite{Bruch,Kar06,Adhikari,Nielsen,Leyronas} at $q \simeq 1.27 \times q_{\rm dim}$ and $q \simeq 16.51 \times q_{\rm dim}$.
\bibitem{Caution} We recall that we have supposed that the interaction between atoms of mass $m_i$ is negligible. We also neglect here the existence of deep bound states of spatial extensions of the order of the range of interatomic forces which are thus very small with respect to $a_z$.
\end{thebibliography}
\end{document}